\documentclass[%
 reprint, 
 amsmath,amssymb, aps, 
]{revtex4-2}

\usepackage{graphicx}
\usepackage{dcolumn}
\usepackage{bm}

\usepackage{wrapfig}
\usepackage{amsfonts}
\usepackage{amsmath}
\usepackage{hyperref}
\usepackage{amssymb}
\usepackage[english]{babel}
\usepackage{graphicx}
\usepackage{epsfig}
\usepackage{bm}
\usepackage{longtable}
\usepackage{verbatim}
\usepackage{longtable}
\usepackage{color}
\usepackage{subfigure}
\usepackage[utf8]{inputenc}
\usepackage{xcolor}
\usepackage{xspace}
\usepackage{ulem}
\usepackage{wasysym}
\usepackage[toc,page]{appendix}
\usepackage{mathtools}
\usepackage{multirow,rotating}
\usepackage{stackrel}
\usepackage{cancel}
\usepackage{pifont}
\usepackage{dcolumn}
\usepackage[compat=1.1.0]{tikz-feynman}
\usepackage{feynmf}
\newcommand{\mathsym}[1]{{}}

\newcommand{\Tr}{\mathop{\rm Tr}\nolimits}

\newcommand{\I}{\ensuremath{\mathrm{i}}}

\newcommand{\sles}[1]{#1 \hspace{-4.5 pt}/}

\newcommand{\keV}{\ensuremath{\mathrm{\,keV}}}
\newcommand{\MeV}{\ensuremath{\mathrm{\,MeV}}}

\newcommand{\eV}{\ensuremath{\mathrm{\,eV}}}

\begin{document}

\preprint{APS/123-QED}

\title{Towards interferometry of neutrino electromagnetism}

\author{Mariia Petropavlova}
 \email{Mariia.Petropavlova@cvut.cz}
 \affiliation{%
Institute of Experimental and Applied Physics, Czech Technical University in Prague, Prague, Czech Republic }
\affiliation{Faculty of Mathematics and Physics, Charles University in Prague, Prague, Czech Republic}
\author{Adam Smetana}%
 \email{Adam.Smetana@cvut.cz}
\affiliation{%
Institute of Experimental and Applied Physics, Czech Technical University in Prague, Prague, Czech Republic 
}%

\date{\today}

\begin{abstract}
    It is predicted within the Standard Model of elementary particles that asymmetric neutrino environments cause rotation of linear polarization of electromagnetic wave -- the birefringence. We demonstrate that this effect is strongly enhanced if additionally the photon is propagating through refractive medium, which effectively increases the photon exposure to the neutrino medium. Our estimate for infrared laser beam in $1\,\mathrm{m}$ long optical fiber exposed to reactor anti-neutrino flux results in linear polarization rotation by the angle $\sim4.6\times10^{-39}\,\mathrm{rad}$. We also derive the proper dependence of the effect on the angle between the directions of photon and neutrino propagation in the laboratory frame. For that purpose we derive the correct form of the basis of polarization four-vectors, which differs from the one widely used in literature. We also estimate the sub-leading optical effect of the neutrino medium due to the neutrino dipole magnetic moment, in terms of a variation of the refractive index and its angular dependence. A rough monochromatic approximation points towards the existence of a resonant enhancement of the effect. 
\end{abstract}

\maketitle

\section{Introduction}


\begin{figure*}[!]
    \begin{minipage}[h!]{0.32\linewidth}
        \center
        \begin{tikzpicture}
         \begin{feynman}
            \vertex (a1);
            \vertex [right=1cm of a1](a2);
            \vertex [left=1.1cm of a1] (a){\(\gamma\)};
            \vertex [right=1.1cm of a2] (a3){\(\gamma\)};
            \vertex [below=1cm of a1] (b1);
            \vertex [below=1cm of a2] (b2);
            \vertex [left=1.1cm of b1] (b){\(\nu\)};
            \vertex [right=1.1cm of b2] (b3){\(\nu\)};
            
    \diagram* { [layered layout, horizontal=a to a3] {
      (a) -- [boson](a1)-- [edge label={\(e\)}](a2)  -- [boson] (a3) ,
      (a1)--[edge label'={\(e\)}] (b1),  (a2)-- [edge label={\(e\)}] (b2),
     (b)  --  (b1) -- [boson, edge label'={\(W\)}] (b2) --  (b3), }
    };
            \end{feynman}
        \end{tikzpicture}
    \end{minipage}
    \hfill
    \begin{minipage}[h!]{0.32\linewidth}
        \center
        \begin{tikzpicture}
         \begin{feynman}
             \vertex (a1);
            \vertex [right=1cm of a1](a2);
            \vertex [above left=0.7cm of a1] (a);
            \vertex [left=0.7cm of a] (c1) {\(\gamma\)};
;
            \vertex [above right=0.7cm of a2] (a3);
            \vertex [right=0.7cm of a3] (c2) {\(\gamma\)};
            \vertex [below left=0.7cm of a1] (b1);
            \vertex [below right=0.7cm of a2] (b2);
            \vertex [left=0.7cm of b1] (b) {\(\nu\)};
            \vertex [right=0.7cm of b2] (b3){\(\nu\)};
            
    \diagram* { [layered layout, horizontal=a to a3] {
      (c1)--[boson](a) -- [edge label=\small{\(e\)}](a1)-- [edge label'=\small{\(\nu\)}](a2)--[edge label=\small{\(e\)}](a3)--[boson](c2) ,
      (a1)-- [boson, edge label=\small{\(W\)}](b1)--(b),  (a2)--[boson, edge label'=\small{\(W\)}](b2) -- (b3),
      (a)--[edge label'=\small{\(e\)}](b1), (a3)--[edge label=\small{\(e\)}](b2),}
    };
            \end{feynman}
        \end{tikzpicture}
    \end{minipage}
    \hfill
    \begin{minipage}[h!]{0.32\linewidth}
        \center
        \begin{tikzpicture}
        \tikzfeynmanset{every blob={/tikz/pattern color={gray!10}}};
         \begin{feynman}
            \vertex (a1);
            \vertex [right=1cm of a1](a2);
            
            \vertex [above=0.6cm of a1] (a);
            \vertex [left=1.2cm of a] (c1) {\(\gamma\)};
;
            \vertex [above=0.6cm of a2] (a3);
            \vertex [right=1.2cm of a3] (c2) {\(\gamma\)};
            
            \vertex [below=0.6cm of a1] (b1);
            \vertex [left=1.2cm of b1] (b) {\(\nu\)};
            
            \vertex [below=0.6cm of a2] (b2);
            \vertex [right=1.2cm of b2] (b3){\(\nu\)};
            \draw[fill=black] (a1) circle(0.7mm); 
            \draw[fill=black] (a2) circle(0.7mm); 
             \vertex[below=0.1cm of a2](l2){$\Lambda_{\nu}$};
            \vertex[below=0.1cm of a1](l1){$\Lambda_{\mu}$};
        
    \diagram* { [layered layout, horizontal=a to a3] {
      (c1)--[boson](a1)-- [edge label'=\small{\(\nu\)}](a2)--[boson](c2) ,
      (a1)-- (b),  (a2)-- (b3),
      }
    };
            \end{feynman}
        \end{tikzpicture}
    \end{minipage}
    \begin{minipage}[h]{1\linewidth}
     {\textcolor{white}.} 
    \end{minipage}
    \begin{minipage}[h]{1\linewidth}
        \begin{tabular}{p{0.32\linewidth}p{0.32\linewidth}p{0.32\linewidth}}
        \centering
        a)  & \centering b)  & \centering c) \\
        \end{tabular}
    \end{minipage}
    \vspace*{-0.3cm}
\caption{ Feynman diagrams for neutrino--photon forward scattering a) one-particle irreducible at one loop level within SM providing the dominant contribution to the photon polarization tensor which is parity-violating, b) one-particle reducible at two loop level within SM providing the sub-dominant contribution to the photon polarization tensor, but providing one of the leading parity-conserving contributions, c) tree level within effective theory of neutrino electrodynamics.}
\label{Fig1}
\end{figure*}
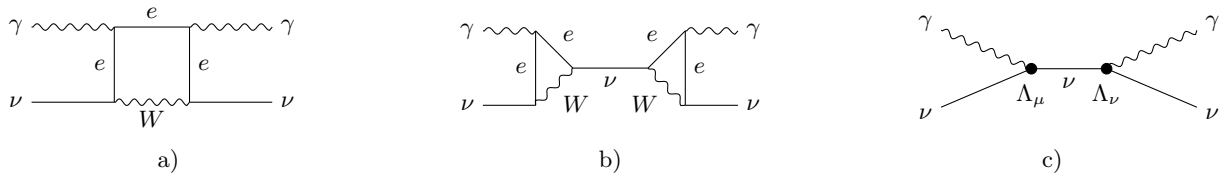

The Standard Model of elementary particles extended for neutrino masses (SM) predicts neutrinos, despite of being neutral fermions, to exhibit electromagnetic properties. They are classified as the charge radius, the electric dipole moment, the magnetic dipole moment, and the anapole moment \cite{Broggini:2012df}. They are result of the screening by quantum fluctuations of neutrinos into virtual charged particles, mainly electrons and $W$-bosons, which are then felt by a photon field. On top of that, the SM predicts a two-photon-fields interaction with neutrinos described effectively by their Rayleigh operator generated at one-loop level \cite{Royer1968,Dicus1993,Mohanty1998,Abbasabadi2001,Abbasabadi2003,Karl:2004bt}.

The experimentally confirmed existence of the neutrino electromagnetic properties, would at first place be an independent probe of the SM itself. Second, it would affect whole areas of the astroparticle physics and cosmology, like, e.g., models of dynamics of compact objects, models of the evolution of the Universe, or the properties of the cosmic neutrino relics. Third, it would open a new channel of neutrino detection.

It is a long-standing exercise to estimate the effect of a medium consisting of neutrino (anti-)particles with nonzero electromagnetic properties on the photon propagation \cite{Royer1968,Nieves1989,Dicus1993,Mohanty1998,Abbasabadi2001,Abbasabadi2003,Karl:2004bt,Nieves2005,Dvornikov2014,Pal2020,Dvornikov2021}. It is a necessary condition that asymmetric neutrino medium, with a non-vanishing difference of neutrino and anti-neutrino densities, $n_\nu-\bar{n}_\nu$, must be assumed in order to avoid complete cancellation of the effect from neutrino and anti-neutrino components. The estimates are based on analysing the forward scattering of photons on neutrinos, for which two classes of diagrams are assumed. The first class consists of one-particle irreducible diagrams of one-loop order in SM, which underlies the Rayleigh operator. One of the examples is on the picture \ref{Fig1}a). The second class consists of one-particle reducible diagrams which are of two-loop order in SM, each loop being responsible for one of two insertions of the electromagnetic form-factors. This class is represented by the diagram on the picture \ref{Fig1}b).  

A significance of individual contributions to the forward-scattering amplitude may be evaluated by counting the power $N$ of the suppression factor $\propto1/M^N_W$ \cite{Mohanty1998}. Here, the leading contribution $N=2$ comes from the leading part of the diagram \ref{Fig1}a), which is purely parity violating. It is responsible for the birefringence, or the optical activity, or the rotary power of the asymmetric neutrino medium, which, e.g., rotates the direction of the linear polarization of the electromagnetic wave. In other words, the left- and right-handed polarizations develop different dispersion relations, and consequently a difference in their phase velocity. This optical activity is experimentally meaningful exclusively for the cases when the polarization of the electromagnetic wave is under control. Any kind of averaging over the polarizations of the electromagnetic wave beam causes cancelling of the parity violating effects. In those cases, the parity-conserving contributions, which are of the sub-leading order, $N=4$, gives the dominant effects, which are exhibited by the modification of the transverse component of the photon polarization tensor, and by the creation of the longitudinal component of the photon polarization tensor. These effects can be in principle observed as the variation of the value of the index of refraction.

The rotary power of the cosmic neutrino background (CNB) has been estimated in some of the previous works with hopelessly small results. The case of a pure CNB medium was initially addressed in \cite{Royer1968} and later elaborated in greater detail in \cite{Abbasabadi2001,Abbasabadi2003}. The result for optical photon energies and for the neutrino--anti-neutrino asymmetry allowed by cosmological models is $\sim47$ orders of magnitude below the rotary power of the intergalactic magnetic fields (IGMF), see the comparative graph in \cite{Dvornikov2021}. A strong enhancement has been identified in \cite{Mohanty1998} as the effect of including additionally an electron component of the intergalactic plasma into the medium, lifting the estimate to $\sim35$ orders of magnitude below the IGMF. Very recently, the authors of \cite{Dvornikov2021} re-estimated the intergalactic plasma enhancement by including properly thermally averaged three-body photon--neutrino--electron scattering and encountered an enhancement by yet another $\sim3$ orders of magnitude, still far below any observability. And yet, we are learning a clear lesson that `slowing-down' of photons by letting them pass through the additional component of a medium increases their exposure to the neutrinos, enhancing the resulting neutrino effect on the photon propagation significantly. Therefore it is relevant to investigate the effect of an optically much denser media.  

In this work we revise the birefringence effect due to the presence of the neutrino asymmetric component of the background medium. We investigate the environment of a refractive medium, e.g., a solid transparent (dielectric) material, pervaded by the neutrino flux. Our goal is to investigate the principal observability of the electromagnetic wave phase shifts due to neutrinos by means of interferometric experiment. Analogous situation of a neutrino medium streaming through plasma with electron component at rest has been analyzed, e.g., in \cite{Nieves2005}. Crucial aspect of our analysis is to correctly identify the directional dependence of the estimated effect given by the angle between photon and neutrino fluxes. In principle, such angular variations may help in detecting these subtle effects. In the last part of this work we also discus the sub-dominant optical effect of parity conserving variation of the refractive index. 

\section{Photon polarizations}

In order to make our analysis clear and directly applicable to any of conceivable laboratory experiments, we want to stick with the calculation in the rest frame of the refractive medium, characterized by the refractive index $n$, typically $1<n<4$, in which electromagnetic wave is characterized by the four-momentum
\begin{equation}\label{k}
    k=(\omega,\mathbf{k})=\omega(1,n\mathbf{\hat{k}})\,,\ \ k^2=\omega^2(1-n^2)\,,
\end{equation}
where $\omega$ is the electromagnetic wave frequency, $\mathbf{k}$ is its three-momentum in the optically dense refractive medium. The unit three-vector $\mathbf{\hat{k}}$ defines the spatial direction of the electromagnetic wave propagation. In the rest frame of the refractive medium, the neutrino medium is in general not in rest. It is rather characterized by the four-velocity
\begin{equation}\label{u}
    u=(u_0,\mathbf{u})=\gamma(1,\beta\mathbf{\hat{u}})\,,\ \ u^2=1\,,
\end{equation}
where $\beta$ is the speed of the neutrino medium and $\gamma$ is the corresponding Lorentz factor. The unit three-vector $\mathbf{\hat{u}}$ defines the spatial direction of the neutrino flux. The direction of the electromagnetic wave and direction of the neutrino flux have a general mutual angle $\theta_{ku}$,
\begin{equation}
    \cos\theta_{ku}=\mathbf{\hat{k}}\cdot\mathbf{\hat{u}}\,.
\end{equation}
As the actual neutrino density experienced by the photon beam depends on the angle $\theta_{ku}$, also the resulting optical effect of the neutrino environment should vary with $\theta_{ku}$. This angular dependence is actually carried by the Lorentz invariant $K$, introduced in former calculations \cite{Nieves1989,Mohanty1998} and others, interpreted as the magnitude of the photon momentum in the rest frame of the medium. 

The form of the Lorentz invariant $K$ has its roots in the proper choice of the orthogonal basis of the photon polarization four-vectors. Let us denote, as usually, three polarization four-vectors as $e_i$, $i=1,2,3$. They satisfy the Lorenz condition 
\begin{equation}\label{Lorenz}
    k_\mu e^\mu_i=0\,.
\end{equation}
Let us choose the first two of the polarization four-vectors, $e_1$ and $e_2$, to be the basis of a two-dimensional sub-space of transverse polarizations 
\begin{equation}\label{e}
    e_{1,2}=(e^0_{1,2},\mathbf{e}_{1,2})=(0,\mathbf{\hat{e}}_{1,2})\,,\ \ e_{1,2}^2=-1\,.
\end{equation}
The second equality is given by a suitable choice of basis, $e^0_{1,2}=0$, which makes it explicit that the transverse polarization four-vectors $e_{1,2}$ are one-to-one connected with the polarization unit three-vectors $\mathbf{\hat{e}}$ which are transverse to $\mathbf{\hat{k}}$,
\begin{equation}\label{ke}
    \mathbf{\hat{k}}\cdot\mathbf{\hat{e}}=0\,.
\end{equation}
The third polarization four-vector $e_3$ corresponds to the longitudinal polarization. These three polarizations together with 
\begin{equation}\label{e0}
e^\mu_0\equiv \frac{k^\mu}{\sqrt{k^2}}
\end{equation}
form a complete orthogonal basis if
\begin{equation}\label{ortho}
    e_\rho\cdot e_\sigma=g_{\rho\sigma}\,,
\end{equation}
which embeds the orthogonality \eqref{ke} and also 
\begin{equation}\label{e1e2}
    \mathbf{\hat{e}}_1\cdot\mathbf{\hat{e}}_2=0\,.
\end{equation}
Given $e^\mu_0$, $e^\mu_1$ and $e^\mu_2$ by equations \eqref{k} and \eqref{e}, the orthogonality condition \eqref{ortho} determines the four-vector $e_3$ to be
\begin{equation}
    e_3^\mu=\frac{\tilde{u}^\mu}{\sqrt{-\tilde{u}^2}}\,,
\end{equation}
where, for the arbitrary neutrino four-velocity $u$,
\begin{equation}\label{tu}
    \tilde{u}^\mu=u^\mu-\frac{(k\cdot u)}{k^2}k^\mu-\frac{(e_1\cdot u)}{e_1^2}e^\mu_1-\frac{(e_2\cdot u)}{e_2^2}e^\mu_2\,.
\end{equation}
The third and fourth terms on the right hand side are missing in \cite{Nieves1989} and all other consequent literature. It leads to the proper angular dependence of the Lorentz invariant $K$ different from the existing literature,
\begin{equation}\label{K4}
    K=\sqrt{(k.u)^2-k^2(1+(e_1\cdot u)^2+(e_2\cdot u)^2)}\,.
\end{equation}
The dependence of $K$ on $(e_{1,2}\cdot u)$ is new. Taking the definitions of $e_{1,2}^\mu$, \eqref{e}, of $k^\mu$, \eqref{k}, and of $u^\mu$, \eqref{u}, we can write the dot products explicitly
\begin{eqnarray}
(k.u) &=& \omega\gamma(1-n\beta\cos\theta_{ku})\,,\\
(e_1.u) &=& -\gamma\beta\sin\theta_{ku}\cos\phi_{eu}\,,\\
(e_2.u) &=& -\gamma\beta\sin\theta_{ku}\sin\phi_{eu}\,,
\end{eqnarray}
where $\phi_{eu}$ is the polar angle of $\mathbf{\hat{e}}_1$ within the plane perpendicular to $\mathbf{\hat{k}}$ relative to the projection of $\mathbf{\hat{u}}$ into this plane. Applying this into \eqref{K4} we come to very simple expression for $K$,
\begin{equation}\label{K}
    K=\omega\gamma|n-\beta\cos\theta_{ku}|\,.
\end{equation}
There are two immediate observations. First, $K$ is always real and non-negative. Second, it can vanish only for $n<1$, in the case of parallel $\mathbf{\hat{k}}$ and $\mathbf{\hat{u}}$ it vanishes just when $\beta=n$ that results in parallel four-vectors $k^\mu$ and $u^\mu$, just see \eqref{k} and \eqref{u}. This is exactly expected from the tensor structure of the parity violation polarization tensor part $\Pi_{P\mu\nu}\propto\varepsilon_{\mu\nu\alpha\beta}k^\alpha u^\beta$ \cite{Mohanty1998}.
This is to be compared with the previous expression for $K$ from, e.g., \cite{Nieves1989,Mohanty1998}, let us denote it by $K'$,
\begin{equation}\label{Kold}
    K'=\sqrt{(k.u)^2-k^2}=\sqrt{K^2-(1-n^2)\omega^2}\,,
\end{equation}
which does not exhibit neither of the above properties. There exists physically reasonable choice of $n$, $\beta$ and $\theta_{ku}$ for which $K'$ is even imaginary. For parallel $\mathbf{\hat{k}}$ and $\mathbf{\hat{u}}$ the requirement for vanishing of $K'$ leads to $\beta=n/\sqrt{n^2-1}$, being meaningful only for $n>0$, which in no way leads to parallel four-vectors $k^\mu$ and $u^\mu$. The expression for $K$ and $K'$ coincide in a special case of light-like photon, $k^2=0$, or in the case of having no preferred inertial frame other than the rest frame of neutrino medium, in which case one can choose $u=(1,0,0,0)$ and the new terms $\propto(e_{1,2}.u)^2$ vanish.

Having chosen the proper basis of polarization four-vectors $e_i^\mu$, $i=1,2,3$, there exists a set of corresponding projectors onto them, these are the transverse projector, $P^{\mu\nu}_\mathrm{T}$, onto transverse modes given by $e_{1,2}^\mu$, and the longitudinal projector, $P^{\mu\nu}_\mathrm{L}$, onto longitudinal mode given by $e_{3}^\mu$,
\begin{eqnarray}
    &&P^{\mu}_{\mathrm{T}\nu}e_{1,2}^\nu=e_{1,2}^\mu\,,\ \ P^{\mu}_{\mathrm{T}\nu}e_{3}^\nu=0\,,\\
    &&P^{\mu}_{\mathrm{L}\nu}e_{3}^\nu=e_{3}^\mu\,,\ \ \ \, \ \ P^{\mu}_{\mathrm{L}\nu}e_{1,2}^\nu=0\,,
\end{eqnarray}
On top of that, there is a parity-violating generator of the transverse polarization rotation, $P^{\mu\nu}_\mathrm{P}$,
\begin{equation}
    P^{\mu}_{\mathrm{P}\nu}e_{1,2}^\nu=\pm\mathrm{i}e_{2,1}^\mu\,.
\end{equation}
These tensors have the explicit form given, e.g., in \cite{Nieves1989},
\begin{eqnarray}
    P^{\mu\nu}_\mathrm{L}&=&\frac{\tilde{u}^\mu\tilde{u}^\nu}{\tilde{u}^2}\,,\\ 
    P^{\mu\nu}_\mathrm{T}&=&g^{\mu\nu}-\frac{k^\mu k^\nu}{k^2}-\frac{\tilde{u}^\mu\tilde{u}^\nu}{\tilde{u}^2}\,,\\
    P^{\mu\nu}_\mathrm{P}&=&\frac{\mathrm{i}}{K}\varepsilon^{\mu\nu\alpha\beta}k_\alpha\tilde{u}_\beta \,,
\end{eqnarray}
keeping in mind that $\tilde{u}^\mu$ and $K$ are defined according to \eqref{tu} and \eqref{K}. From this point on, the analysis follows according to formerly established lines described, e.g., in \cite{Nieves1989,Mohanty1998}. The photon polarization tensor can be decomposed as
\begin{equation}
    \Pi^{\mu\nu}=\Pi_\mathrm{T} P^{\mu\nu}_\mathrm{T} + \Pi_\mathrm{L} P^{\mu\nu}_\mathrm{L} + \Pi_\mathrm{P} P^{\mu\nu}_\mathrm{P}\,.
\end{equation}
It is the parity-violating polarization function $\Pi_\mathrm{P}$ which is responsible for the birefringence effect.

\section{Birefringence}

The leading term in $1/M^2_W$ expansion of the neutrino induced correction to the polarization tensor $\Pi^{\mu\nu}(k)$ comes from the one-loop diagram Fig.\ref{Fig1}a) and contributes exclusively to the parity-violating polarization function $\Pi_\mathrm{P}(k^2,K)$ responsible for the birefringence effect. The polarization function $\Pi_\mathrm{P}(k^2,K)$ has the form \cite{Mohanty1998} under assumption $k^2\ll m^2_e\ll M^2_W$
\begin{equation}\label{PiP}
    \Pi_\mathrm{P}(k^2,K)=\frac{\sqrt{2}G_F\alpha}{3\pi}\frac{k^2}{m^2_e}(n_\nu-n_{\bar\nu})K\,,
\end{equation}
again keeping in mind the definition of $K$ \eqref{K}. The quantities $n_{\nu(\bar{\nu})}$ are the (anti-)neutrino medium rest frame densities.

As shown in \cite{Mohanty1998}, from diagonalization of the photon full propagator, which includes both the refractive medium effect in the form of $\Pi_T\sim k^2=\omega^2(1-n^2)$, and the neutrino medium effect in the form of $\Pi_P$ given by \eqref{PiP}, the dispersion formula for the two transverse modes has the form
\begin{equation}
    k^2_\pm\equiv\omega^2-|\mathbf{k}_\pm|^2=\Pi_T\pm\Pi_P\,.
\end{equation}
Here we have introduced momenta $|\mathbf{k}_\pm|$, which represents magnitude of the actual photon wave vectors that includes both refractive and neutrino medium effects. 
The rotary power as the linear polarization rotation angle per unit length is then expressed in terms of the actual photon momenta $\mathbf{k}_\pm$ as
\begin{equation}\label{rotdef}
    \frac{\phi}{l}=\frac{1}{2}(|\mathbf{k}_-| - |\mathbf{k}_+|)\simeq\frac{\Pi_P}{2\omega n}\,.
\end{equation}
After combining all the pieces, \eqref{k}, \eqref{K}, \eqref{PiP}, into the expression \eqref{rotdef} for rotary power we are getting our key result
\begin{equation}\label{rot}
    \frac{\phi}{l}=\frac{G_F\alpha}{3\sqrt{2}\pi}\frac{\omega^2(1-n^2)}{m^2_e}\gamma(n_\nu-n_{\bar\nu})|1-\frac{\beta}{n}\cos\theta_{ku}|\,.
\end{equation}

Using the formula \eqref{rot}, we can make a numerical estimate of the effect in the conceivable physical situation. A linearly polarized infrared laser beam of $\omega\sim1\eV$ would pass through a silicon optical fiber with the refractive index $n\sim3.5$ in the vicinity of a nuclear reactor approximated as a point-like source of anti-neutrino flux of $f=10^{21}\,\mathrm{s}^{-1}$ (here we take into account also slow anti-neutrinos). The $l=1\,\mathrm{m}$ long optical fiber is placed to a $d=5\,\mathrm{m}$ distance from the reactor core in the perpendicular direction to the anti-neutrino flux, so $\cos\theta_{ku}=0$. The local density of antineutrinos in the optical fiber is then $\gamma(n_\nu-n_{\bar\nu})=f/(4\pi c d^2)\approx10^4\,\mathrm{cm}^{-3}$ and the resulting angle, by which the linear polarization of the laser beam is rotated, is $\phi\sim4.6\times10^{-39}\,\mathrm{rad}$. This appears to be undetectably small quantity. On the other hand it should be compared with the previous results, \cite{Dvornikov2021}, where a comparable magnitude of the same effect has been achieved by letting pass the astronomical photons through the intergalactic void filled with plasma electrons and relic neutrino medium over the distance of a Hubble radius $l=l_H\sim10^{26}\mathrm{m}$. Notice in \eqref{rot} the dependence of the effect on the refractive index factor $(1-n^2)\sim\mathcal{O}(1-10)$ which is by $26$ orders of magnitude larger then the corresponding quantity $\omega_p^2/\omega^2$ describing the electron component of the intergalactic plasma characterized by the plasma frequency $\omega_p\sim10^{-13}\eV$. It is the optical density that compensates the need for Hubble distances down to laboratory scales to get the effect of the same magnitude. 


One can think of increasing the effect by some orders of magnitude via changing some of the parameters. First, the effect is linear in the local neutrino density. Approaching closer towards the core of the nuclear reactor and increasing its power would help, conceivably, each by one order of magnitude, for the price of exposing the experimental setup to an extreme thermal and radiation environment. Second, the effect is linear in the optical fiber length. It is conceivable to design a $l=\mathcal{O}(100\,\mathrm{km})$ optical fiber. However, the optical stability of the optical fiber against thermal and vibration noise at the required level is extremely sensitive to its length, already $l=1\,\mathrm{m}$ is beyond current technological capabilities. A double-link laser setup using two or more laser sources of different wavelengths simultaneously in the same optical fiber would provide a set of experimental data, whose linear combination might suppress the noise while keep the signal, provided that the noise and signal have distinct dispersion. For increasing the effective optical fibre length, unfortunately, the fibre-based Fabry-Pérot cavity, which can significantly enhance the effective photon path without increasing the noise, does not bring any advantage, because the sign of the effect, i.e., of the angle $\phi$, is given by the sense of the photon propagation so that the reflected laser beam would get exactly opposite linear polarization rotation. Third, the effect is quadratic in $\omega$ of the electromagnetic wave. Using harder photons, e.g., X-rays with $\omega=1\keV-100\keV$ may bring factor $10^6 - 10^{10}$ however correspondingly the refractive index drops to unity with inverse squared $\omega$ so that $1-n^2\propto\omega^{-2}$ which consequently cancels out the advantage coming from the shorter wavelength. On top of that for harder photons the attenuation increases. 

\section{Variation of refractive index}

In this last part of our work, we would like to address briefly the sub-dominant effect of the order of $1/M^4_W$ that becomes the leading effect in the situation, when the birefringence discussed above is averaged away for some reason such as the laser beam is not polarized, or the polarization is destroyed along the propagation due to the imperfection of the optical fiber, or the Fabry-Pérot cavity resonator is used. 

There are many Feynman diagrams contributing to the forward scattering at the $1/M^4_W$ order. Two of them are depicted in the Fig.\ref{Fig1}a) and b). We however discuss the diagram Fig.\ref{Fig1}c), coming from the effective neutrino electromagnetic theory \cite{Broggini:2012df}, which is induced by nonzero neutrino electromagnetic coupling $\Lambda_\mu$, which corresponds, e.g., to the neutrino dipole magnetic moment. In SM the dipole magnetic moment of Dirac neutrino is predicted from the vertex loop in Fig.\ref{Fig1}b) to be $\propto1/M^2_W$ and among other neutrino electromagnetic couplings to be the strongest one, though with its value $\mu_\nu\sim10^{-19}\mu_\mathrm{B}$ it is still $\sim7$ orders of magnitude below current experimental capabilities. 

Following the analogous calculation of \cite{Mohanty1998}, we calculate the part of the photon polarization tensor induced by the electromagnetic coupling, $\Pi^{(\Lambda)}_{\mu\nu}$ and consider only the medium-induced part that we indicate by the prime, $\Pi^{(\Lambda)'}_{\mu\nu}$,
\begin{equation}\label{PiL}
    \Pi^{(\Lambda)'}_{\mu\nu}(k)=\I\int\frac{\mathrm{d}^4p}{(2\pi)^4}\Tr\{S'(p,u)\left[T_{\mu\nu}(p,k)+T_{\nu\mu}(p,-k)\right]\}\,,
\end{equation}
where
\begin{eqnarray}
    S'(p,u)&=& (\sles{p}+m_\nu)2\pi\I\delta(p^2-m^2_\nu)\\
    &&\times\left[\theta(p\cdot u)f(p\cdot u)+\theta(-p\cdot u)\bar{f}(-p\cdot u)\right] \,,\nonumber\\
    T_{\mu\nu}(p,k)&=&\frac{\Lambda_\mu(\sles{p}+\sles{k}+m_\nu)\Lambda_\nu}{(p+k)^2-m^2_\nu} \,,
\end{eqnarray}
where $f$ and $\bar{f}$ are the distribution functions of neutrinos and anti-neutrinos, respectively. The integral \eqref{PiL} can be conveniently calculated in the neutrino medium rest-frame by Lorentz transforming it first, so that $p\cdot u\rightarrow p_0$ and $k\rightarrow k_u$ becomes the photon four-momentum in the neutrino medium rest frame. After the integration we Lorentz transform the result back to the laboratory frame. In what follows we keep in mind the physical setup of the laser beam in an optical fiber exposed to reactor antineutrinos, and simplify significantly our calculation by considering the antineutrino stream as perfectly collinear and monochromatic, specified in the neutrino medium rest frame as
\begin{equation}\label{mono}
    f=0\,,\ \ \ \bar{f}=(2\pi)^3 n_{\bar{\nu}}\delta^{(3)}(\mathbf{p})\,,
\end{equation}
where $n_{\bar{\nu}}$ is the density of anti-neutrino medium in its rest frame.
Then we first perform the $p_0$ integration with the help of $\delta(p^2-m^2_\nu)$ 
\begin{equation}\label{PiL3d}
    \Pi^{(\Lambda)'}_{\mu\nu}(k_u)=\int\frac{\mathrm{d}^3\mathbf{p}}{(2\pi)^3}\frac{\bar{f}}{2|p_0|}\left.\frac{\Tr_{\mu\nu}(k_u)+\Tr_{\nu\mu}(-k_u)}{4(k_u\cdot p)^2-k_u^4}\right|_{p_0=-\epsilon}
\end{equation}
where $\epsilon\equiv\sqrt{\mathbf{p}^2+m^2_\nu}$ and
\begin{equation}
    \Tr_{\mu\nu}(k)=\Tr\{(\sles{p}+m_\nu)\Lambda_\mu(\sles{p}+\sles{k}+m_\nu)\Lambda_\nu\}(k^2-2(k\cdot p))\,,
\end{equation}
Due to the monochromatic approximation \eqref{mono} the three-momentum integration is trivial and basically it comprises of the substitution in the integrand $p^\mu\rightarrow(m_\nu,0,0,0)\equiv m_\nu u^\mu_u$, where $u^\mu_u$ is the anti-neutrino medium four-velocity in its rest frame. Finally we Lorentz transform the result back to the laboratory frame, so that we remove the $u$-subscripts $k_u$ and $u_u$. The final result is  
\begin{equation}
    \Pi^{(\Lambda)'}_{\mu\nu}(k)=\frac{1}{m_\nu}\frac{\Tr_{\mu\nu}(k,u)+\Tr_{\nu\mu}(-k,u)}{4m_\nu^2(k\cdot u)^2-k^4}n_{\bar{\nu}}\,.
\end{equation}
For the coupling of the dipole magnetic moment we have $\Lambda_\mu=\mu_\nu k^\alpha\sigma_{\alpha\mu}$ and the resulting contribution to the photon transverse polarization function is 
\begin{equation}
    \Pi^{(\mu)'}_\mathrm{T}(k)=8\mu_\nu^2 m_\nu k^4\frac{2+(\mathbf{k}\cdot\mathbf{u})^2/|\mathbf{k}|^2}{k^4-4m_\nu^2(k\cdot u)^2}n_{\bar{\nu}}\,.
\end{equation}
Applying this result to the photon propagation in the refractive medium, i.e. using \eqref{k} and \eqref{u}, we can make the estimate of the variation of the index of refraction due to the neutrino medium
\begin{equation}\label{Dn}
    \Delta n\approx\frac{\Pi^{(\mu)'}_\mathrm{T}}{2\omega^2}\simeq-\frac{\mu_\nu^2 \gamma n_{\bar{\nu}}}{E_\nu}(1-n^2)^2 \frac{\sin^2\theta_{ku}}{(1-n\beta\cos\theta_{ku})^2-\Gamma}\,,
\end{equation}
where $\Gamma = \frac{\omega^2(1-n^2)^2}{4E^2_\nu}\sim10^{-12}$ for the reactor anti-neutrino energy $E_\nu=\gamma m\sim\mathcal{O}(\MeV)$ and for the visible light frequency $\omega\sim\mathcal{O}(\eV)$. For these conditions we are getting extremely tiny correction to the refractive index $\Delta n\sim1.2\times10^{-65}\Theta(\theta_{ku})$, where the function $\Theta(\theta_{ku})$ carries the angular dependence which provides typically factor of $\mathcal{O}(1)$. Interestingly, however, the angular factor may be source of a resonant enhancement if the angle $\theta_{ku}$ is tuned to be close to $\cos\theta_{ku}\sim1/(n\beta)$. Similar resonances are typical for the media composed of neutral particles with dipole magnetic moment. The angular factor in the expression \eqref{Dn} due to the assumption of perfect collinearity and monochromaticity of the neutrino flux, could reach an unlimited values, which is of course unphysical consequence of unrealistic approximation. To analyse properly the resonant enhancement is subject of future work. Now we just speculate that the values of the resonant enhancement up to the level $\sim\Gamma^{-1}$ could be potentially accessible if the collinearity and angular stability of the neutrino flux with respect to the photon direction could be tuned down to the level of the angular spread $\Delta\cos\theta_{ku}<\sqrt{\Gamma}$. 

Notice, that the resonance within the angular factor in principle appears in two regimes: The first regime is the ultra-relativistic regime, characterized by $\Gamma\rightarrow0$ and $\beta\rightarrow1$, in which the resonance is achieved by tuning the angle $\theta_{ku}$. The second regime is the non-relativistic regime, characterized by $\Gamma\sim\mathcal{O}(1)$ and $\beta\rightarrow0$, in which the resonance is achieved by tuning the ratio of neutrino and photon $\theta_{ku}$. Interestingly, $\Gamma\sim\mathcal{O}(1)$ for $\omega$ from the visible-light range or similar, is only accessible thanks to lightness of neutrinos so that one can arrange $m_\nu\sim\omega$. If neutrinos were heavier, harder photons would be needed, for the high price of loosing the advantage of the refractive index being $(1-n^2)\sim\mathcal{O}(1)$. Hypothetically, if the CNB were much colder than currently predicted, $T\ll\beta_\odot m_\nu\ll1.95\,\mathrm{K}$, so that the peculiar motion of the Earth with respect to the Universal grid, characterized by the speed $\beta_\odot\sim10^{-4}$, would produce pretty monochromatic neutrino flux through the detector, then one could, at least in principle, tune $\omega$ to probe the resonance. In reality, however, the thermal motion of the CNB neutrinos totally smears out the resonance of this kind.

\section{Conclusions}

In the present work, we have addressed a novel idea of the interferometric detection of neutrino electromagnetic properties in laboratory. For that purpose we have revised the formula describing the birefringence effect for the case of neutrino flux streaming through an ordinary transparent refractive medium, such as an optical fibre. Though the estimate of the magnitude of the effect turned out to be undetectable by current interferometric technologies, the result exhibits a significant reduction of the length scale of the phenomenon down to the laboratory scales, compared to the cosmological scales addressed in the existing literature so far. Additionally, within very rough monochromatic approximation, we have derived and estimate the sub-dominant optical effect, the variation of the refractive index, induced by the neutrino dipole magnetic moment for the same environment. The approximation points towards the existence of a resonant enhancement of the effect, whose advantage would be accessible only under a delicate tuning of the monochromaticity and directional uniformity of the neutrino flux. To estimate the resonant enhancement factor for realistic neutrino fluxes, the analysis beyond the rough approximation used here, is necessary, which is the subject of future work.      

In order to properly derive the angular dependence of the present optical effects, we have revised the description, used in the related literature, of the photon polarizations in medium in general Lorentz frame. We identified a discrepancy from the description used in the existing literature leading to distinct angular dependencies of the effects. We have presented simple arguments in favor of our description based on directional dependence of the birefringence, inferred from general covariant form of the parity-violating photon polarization tensor component.

The results presented in this work suggest the interferometry of neutrino electromagnetism to be still far from current technological capabilities. They are, however, presented here to trigger the interest in this research direction. The existence and nature of the possible resonance certainly deserves further and more thorough theoretical investigation. Sophisticated experimental setups, which shift the interferometric effects further by orders of magnitude, may be revealed by following research. New technological concepts, such as the optical-fibre based atomic interferometry, may shrink the gap between the theory and experiment in foreseeable future. On top of that the present ideas, concepts and results are in principle applicable to the case of a fermionic dipolar candidate for dark matter, see, e.g., \cite{Sigurdson2004}, something that deserves deserves a proper attention.

\section*{Acknowledgements}

We acknowledge the support of the Institute of Experimental and Applied physics, Czech Technical University in Prague, for many years of support. This work was supported by European Regional Development Fund-Project Engineering Applications of Microworld Physics (Grant No. CZ.02.1.01/0.0/0.0/16\textunderscore019/0000766). Work of M.P. was additionally supported by the PhD. scholarship of Charles University in Prague. 

\bibliography{references.bib}

\end{document}